\documentclass[12pt]{article}

\usepackage{times, amsfonts, amstext, amsmath, amssymb, amsthm, titlesec}

\usepackage[authoryear]{natbib}

\usepackage[breaklinks, colorlinks=true,allcolors=blue]{hyperref}
\usepackage{etoolbox}

\usepackage{float}

\makeatletter

\patchcmd{\NAT@citex}
{\@citea\NAT@hyper@{%
		\NAT@nmfmt{\NAT@nm}%
		\hyper@natlinkbreak{\NAT@aysep\NAT@spacechar}{\@citeb\@extra@b@citeb}%
		\NAT@date}}
{\@citea\NAT@nmfmt{\NAT@nm}%
	\NAT@aysep\NAT@spacechar\NAT@hyper@{\NAT@date}}{}{}

\patchcmd{\NAT@citex}
{\@citea\NAT@hyper@{%
		\NAT@nmfmt{\NAT@nm}%
		\hyper@natlinkbreak{\NAT@spacechar\NAT@@open\if*#1*\else#1\NAT@spacechar\fi}%
		{\@citeb\@extra@b@citeb}%
		\NAT@date}}
{\@citea\NAT@nmfmt{\NAT@nm}%
	\NAT@spacechar\NAT@@open\if*#1*\else#1\NAT@spacechar\fi\NAT@hyper@{\NAT@date}}
{}{}
\makeatother

\usepackage{tabularx, siunitx, booktabs}
\usepackage[labelfont=bf, labelsep=period, singlelinecheck=false]{caption}

\long\def\symbolfootnote[#1]#2{\begingroup%
\def\thefootnote{\fnsymbol{footnote}}\footnote[#1]{#2}\endgroup}

\titleformat{\section}{\large\bfseries}{\thesection.}{.5em}{}
\titlespacing*{\section}{0pt}{*3}{*2}
\titleformat{\subsection}{\normalfont\bfseries}{\thesubsection.}{.5em}{}
\titlespacing*{\subsection} {0pt}{*3}{*2}
\titleformat{\subsubsection}{\normalfont\bfseries}{\thesubsubsection.}{.5em}{}
\titlespacing*{\subsubsection} {0pt}{*3}{*2}

\theoremstyle{plain}

\numberwithin{equation}{section}

\setcitestyle{aysep={}}

\begin{document}

\title{\textbf{\LARGE Two-Stage and Sequential Unbiased Estimation of $N$ in Binomial Trials, when the Probability of Success $p$ is Unknown}}

\date{}

\begingroup
\let\center\flushleft
\let\endcenter\endflushleft
\maketitle
\endgroup

\author{
\vskip -1cm
\noindent
{\large Yaakov Malinovsky\textsuperscript{a} and Shelemyahu Zacks\textsuperscript{b}}

\noindent \textsuperscript{a}Department of Mathematics and Statistics, University of Maryland, Baltimore County, Baltimore, Maryland, USA

\noindent \textsuperscript{b} Department of Mathematical Sciences, Binghamton University,
Binghamton, New York, USA}

\symbolfootnote[0]{\normalsize \flushleft \textbf{CONTACT} Yaakov Malinovsky
Department of Mathematics and Statistics, University of Maryland, Baltimore County, 1000 Hilltop Circle, Baltimore,
MD 21250, USA; \href{mailto:yaakovm@umbc.edu}{yaakovm@umbc.edu}}
\bigskip

{\small \noindent\textbf{Abstract:}
We propose two--stage and sequential procedures to estimate the unknown parameter $N$ of a binomial distribution with unknown parameter $p$, when we reinforce data with an independent sample of a negative--binomial experiment having the same $p$.

\section{Introduction}

Estimating the number $N$ of Bernoulli trials when the only available information is the number of successes among those trials is a classical problem on which many papers are available in the literature. For example, see \cite{W1914}, \cite{B1953}, \cite{BD1981}, \cite{DR2005}, \cite{SHV2021}, and others.
The motivation for considering such a problem can be found in ecological problems of estimating the size of a population, in fishery, in software reliability, and in more areas besides.
Existing methodologies are not satisfactory with respect to the accuracy and precision of the suggested estimators. Furthermore, based on one random sample from a binomial experiment, when $p$ is unknown, an unbiased estimator of $N$ does not exist \citep{DR2005}. In the present paper we are following a recent result of Rasul Khan \citep{K2021}, who obtained an unbiased estimator of $N$ by reinforcing the data with an independent sample of a negative--binomial experiment having the same parameter $p$. Two--stage and sequential sampling for estimating $N$ with a prescribed precision requirement, when the probability of success $p$ is known was published a few years ago by \cite{DZ2016}.

To be more specific, let $X$ denote a random variable having a binomial distribution, $B(N,p)$. Let $T$ be a random variable having a Pascal distribution, $Pasc(m,p)$, whose p.m.f. presents the probability of the number of Bernoulli trials required to observe $m$ successes. We assume that $X$ and $T$ are independent, having the same $p$ parameter. It is well known that $E(X)=Np$ and $E(T)=m/p $, hence $E\left(XT/m\right)=N$. Thus, $\hat{N} =XT/m$ is an unbiased estimator of $N$. We will determine the number of observations on $X$ and $T$ that are required to obtain such an unbiased estimator with a prescribed precision. We will construct confidence intervals around ${\hat{N}}$ whose half width is equal to a certain proportion of $N$.
For this purpose we need two--stage or sequential sampling.

For example, in which our methodology applies directly is the following.
A company is saleing a desired commodity over the internet. The company publicized that the number of
visits of possible clients is between 10,000 to 20,000 a day. A visitor may buy only one product at a time.
Not all visitors buy the product. The actual number  of visitors, $N$, and the probability that a visitor will buy
the product, $p$, are not published. For a specified day, the company agreed to inform that the number of
visitors that bought the product was 3580, and independently the  number of visitors until $m=10$ products were
sold was 25. With this information we can estimate $N$ unbiasedly and $p$.

In Section $2$ we derive the exact distribution of  $\hat{N}$ based on one observation on $X$ and one on $T$.
Three exact moments are presented, as well as the variance, and skewness. A common confidence interval for $N$ is given at the end of this section.

Section $3$ studies an alternative unbiased estimator based on the product of the averages of $k$ $X$ values and $k$ $T$ values. It is shown that for each $k$ greater than 1, the variance of this alternative estimator is smaller than that of the simple average estimator. Section $4$ develops the two--stage and sequential procedures when the parameters $N$ and $p$ are unknown.

\section{Distribution of  $\hat{N}$ and its Moments}

We start with the distribution of $\hat{N}$.
Let $B(j, N ,p) ,j =0,\ldots, N$  denote the c.d.f. of the binomial distribution, and $b(j, n ,p)$ its corresponding p.m.f.
Since $T$ and $X$ are independent random variables, we can write the probability of the event $\left\{\hat{N} >\xi\right\}$ for nonnegative $\xi$ as
\begin{align}
\label{eq:2}
&
P(\hat{N}>\xi)
=\sum_{x =1}^{N}b(x, N ,p)P(T >(\xi/x)m)
                                        =\sum_{x =1}^{N}b(x, N ,p)P(Y >m((\xi/x) -1)),
\end{align}
where $Y=T-m$ is a negative-binomial random variable, with parameters (m, p). With formula \eqref{eq:2}  we can calculate all kind of functionals of the c.d.f. of $\hat{N}$. In the following table we present the survival distribution of $\hat{N}$ for the case of $N=300$, $p=.4$ and $m=10$.


\begin{table}[H]
\caption {Survival probabilities of $\hat{N}$ for $N=300, p=.4$  and $m=10$.}
\label{eq:t1}
$
\begin{array}{cc}
n & P(\hat{N} >n)\\
100 & 0.9999 \\
150 & 0.9954 \\
200 & 0.9263 \\
250 & 0.7212 \\
300 & 0.4500 \\
350 & 0.2350 \\
400 & 0.1022 \\
450 & 0.0401 \\
500 & 0.0141
\end{array}
$
\end{table}

Notice that for every $n=0,1,\ldots$
\begin{equation}
P\left(n<\hat{N}\leq n+1\right)=P\left(\hat{N}>n\right)-P\left(\hat{N}>n+1\right).
\end{equation}

The median of the distribution of example in Table \ref{eq:t1} is 291. This is slightly smaller than the expected value (300) and indicates that the distribution is positively skewed. Moments of this distribution can be calculated numerically with the above function. In the following, we derive analytically some moments of the distribution.
\bigskip

\noindent
{\bf Moments}:
\medskip

Due to independence of $X$ and $T$, the $r$-th moment of $\hat{N}$ is  $E\left\{(XT/m)^{r}\right\} =E(X^{r})E(T^{r})/m^{r}$. Thus, we have to multiply the $r$-th moment of the binomial X, by the $r$-th moment of the Pascal $T$. These moments can be obtained by differentiating the corresponding moment-generating functions \citep{KS1958}.

The moments of $\hat{N}$ are
\begin{align}
&
E(\hat{N})=N,\,\,\,\,
E(\hat{N}^{2})=N(Np - p + 1)(m - p + 1)/(mp),\nonumber
\\
&
E(\hat{N}^{3})=(N(N - 1)(N - 2)p^3 + 3N(N - 1)p^2 + Np)(m^2 - 3mp + 3m + p^2 - 3p + 2)/(m^2 p^3).
\end{align}

It follows that, for $q=1-p$,
\begin{equation}
\label{eq:Var}
Var(\hat{N}) =Nq(Np +q +m)/(pm)
\end{equation}

and the skewness coefficient of the distribution of $\hat{N}$ is
\begin{equation}
Skew(\hat{N})=\frac{E(\hat{N} -N)^3}{Var(\hat{N})^{3/2}}
=\frac{E(\hat{N}^{3})-N^3-3NVar(\hat{N})}{Var(\hat{N})^{3/2}}.
\end{equation}

In the following table we present some characteristics of the distribution of $\hat{N}$. In addition to the $std(\hat{N})=\sqrt{Var(\hat{N})}$,  we present the Skewness coefficient, $Skew(\hat{N})$.

\begin{table}[H]
\caption{Characteristics of the Distribution of $\hat{N}$ .}
$
\begin{array}{cccccc}N & p & m & std(\hat{N}) & Skew(\hat{N})\\
500 & .6 & 10 & 101.719 &  0.7057 \\
500 & .6 & 20 & 73.075 &  0.5012 \\
500 & .6 & 30 & 60.590 &  0.4105 \\
500 & .6 & 40 & 53.260 &  0.3562 \\
500 & .3 & 10 & 136.925 &  0.6837 \\
500 & .3 & 20 & 99.787 &  0.5045 \\
500 & .3 & 30 & 83.829 &  0.4260 \\
500 & .3 & 40 & 74.579 &  0.3789
\end{array}
$
\end{table}

We see in the table above that the distribution is skewed to the right, and the skewness coefficient getting smaller as $m$ grows. The std is decreasing with increasing $m$. This can be explained by the fact that increasing $m$ requires stochastically increasing samples of $T$. To construct confidence intervals of a certain size, one needs large samples, according to the desired size of the intervals.

Let $\left\{X_k, k\geq 1\right\}$, denote a sample of $k$ i.i.d. values of $B(N, p)$. Let $\left\{T_k, k\geq 1\right\}$ denote an sample of $k$ i.i.d. values of Pascal (m, p). $\left\{X_k, k\geq 1\right\}$ and $\left\{T_k, k\geq 1\right\}$ are independent. We consruct a sample of $k$ i.i.d. unbiased estimators $U_{k}=\left\{\hat{N}_1,\ldots,\hat{N}_k\right\}$,
where $\hat{N}_j=X_j T_j/m$. Let
\begin{equation}
\hat{\bar{N}}_k=\frac{1}{k}\sum_{i=1}^{k}\hat{N}_i.
\end{equation}

$\hat{\bar{N}}_k$ is an unbiased and strongly consistent estimator of $N$. Furthermore, for large values of $k$, by central limit theorem (CLT), its distribution is approximately normal, i.e.
\begin{equation}
\label{eq:CLT}
P\left(\hat{\bar{N}}_k\leq t\right)\cong \Phi\left(\frac{\sqrt{k}(t-N)}{\sqrt{Var(\hat{N}_1)}}\right),
\end{equation}
where $Var(\hat{N}_1)$ is given in \eqref{eq:Var}, and $\Phi$ is the distribution function of the standard normal distribution.

An unbiased estimator of $Var(\hat{N}_1)$ is given by
\begin{equation}
S^2_k=\frac{1}{k-1}\sum_{i=1}^{k}\left(\hat{N}_i-\hat{\bar{N}}_k\right)^2.
\end{equation}

We consider the following confidence interval for $N$, with limits
\begin{equation}
\hat{\bar{N}}_k\pm z_{1-\alpha/2}S_{k}/\sqrt{k}.
\end{equation}

The coverage probability, $CP_{k}$, of this interval is
\begin{equation}
CP_{k}=P\left(\hat{\bar{N}}_k - z_{1-\alpha/2}S_{k}/\sqrt{k}\leq N \leq \hat{\bar{N}}_k+ z_{1-\alpha/2}S_{k}/\sqrt{k} \right).
\end{equation}

This probability depends on the parameter $\theta=\left\{N, p, m\right\}$.
Since $S^2_k$ is a strongly consistent estimator of $Var(\hat{N}_1)$, i.e., $S^2_k\rightarrow Var(\hat{N}_1)$ a.s. as $k\rightarrow \infty$, by Slutsky's theorem and \eqref{eq:CLT}
\begin{equation}
CP_{k}=P_{\theta}\left(N - z_{1-\alpha/2}S_{k}/\sqrt{k}\leq \hat{\bar{N}}_k \leq N+ z_{1-\alpha/2}S_{k}/\sqrt{k} \right)
\rightarrow 2\Phi\left(z_{1-\alpha/2}\right)-1=1-\alpha,
\end{equation}
as $k\rightarrow \infty$.

In order to obtain estimates of the coverage probability of this confidence interval, we make originally $2k$ estimates of
$\hat{N}$. We use $k$ estimates to compute $\hat{\bar{N}}_k$, and {\it{independently}} the other $k$ estimates to compute $S_{k}^{2}$ . Since these two estimators
are independent, the conditional coverage probability of the confidence interval is approximately
\begin{equation}
\label{eq:cp}
CP_{k}(S_{k}^{2}) \simeq 2\Phi (z_{1 -\alpha /2}S_k/D) -1,
\end{equation}

where D=$\sqrt{Var(\hat{{N}}_1)}$. Finally, the coverage probability is  $CP_k =E\{CP_k(S_k^{2})\} .$

For example, in the case of $N=100, p=0.6, m=10$, and $\alpha  =0.05,$ we obtain that $Var(\hat{{N}}_1)=469.333$ by simulations we obtained, independently,
$S_{100}^{2} =436.5608.$  According to \eqref{eq:cp} we obtain the conditional coverage probability of $CP_k(436.561) =0.9413.$
The CP can be estimated by the mean of several such conditional CP's.

\section{Alternative Unbiased Estimator}

Suppose we have two independent samples of $k_1$ $X$'s and $k_2$ $T$'s. Consider the estimator, constructed as the product of the two sample means, i.e.
${\overline{X}}_{k_{1}} =\sum _{i =1}^{k_{1}}X_{i}/k_1$  and similarly $\overline{T}_{k_{2}} =\sum _{i =1}^{k_{2}}T_{i_{\,}}/k_{2}$.
Namely,

\begin{equation}
\label{eq:unb}
{\overline{N}}_{k_1, k_2} ={\overline{T}}_{k_{2}}\,{\overline{X}}_{k_{1}}/m =\frac{1}{mk_{1}k_{2}}\sum _{i =1}^{k_1^{\,_{\,_{\,}}}}\sum _{j=1}^{k_2}X_{i}T_{j}
\end{equation}

Since $X$ and $T$ are independent, the estimator \eqref{eq:unb} is unbiased. Moreover, due to independence of $X$ and $T$, the second moment of  ${\overline{N}}_{k_1, k_2}$ is the product of the corresponding second moments, divided by $m^{2}$. The second moments of ${\overline{X}}_{k_{1}}$ and ${\overline{T}}_{k_{2}}$  are,
\begin{equation}
E({\overline{X}}_{k_{1}}^{2}) =Npq/k_1 +(Np)^{2},\,\,\,
E({\overline{T}}_{k_{2}}^{2}) =mq/(k_2p^{2}) +(m/p)^{2}.
\end{equation}
It follows that the variance of $\overline{N}_{k_1, k_2}$ is
\begin{equation}
\label{eq:vv}
Var({\overline{N}}_{k_1, k_2})=\frac{Nq}{mp}\left(\frac{Np}{k_2}+\frac{m}{k_1}+\frac{q}{k_1 k_2}\right) .
\end{equation}

If $k_1=k_2=k$, we obtain
\begin{equation}
\label{eq:vvv}
Var({\overline{N}}_{k})=\frac{Nq}{mpk}\left(Np+m+\frac{q}{k}\right) .
\end{equation}

From equation \eqref{eq:Var} we conclude that the variance of the sample mean ${\displaystyle{\hat{\bar{N}}_k=\frac{1}{k}\sum_{i=1}^{k}\hat{N}_i}}$ is

\begin{equation}
\label{eq:vh}
Var(\hat{\bar{N}}_k) =(Nq/kmp)(Np +m +q).
\end{equation}

It follows that the variance of ${\overline{N}}_{k}$ is smaller than the variance of $\hat{\bar{N}}_k$  for all $k>1$. The efficiency of
$\hat{\bar{N}}_k$ relative to that of $\overline{N}$  is
\begin{equation}
RE =Var({\overline{N}}_{k})/Var(\hat{\bar{N}}_k)=(Np +m +q/k)/(Np +m +q).
\end{equation}
In the following table we present a few numerical example of this relative efficiency

\begin{table}[H]
\caption {Relative Efficiency.}
$
\begin{array}{llllll}N & p & m & k & RE \\
500 & .6 & 10 & 10 & 0.99884 \\
500 & .6 & 10 & 20 & 0.99878 \\
500 & .6 & 10 & 50 & 0.99874 \\
500 & .6 & 10 & 100 & 0.99872 \\
500 & .6 & 10 & 1000 & 0.99871 \\
500 & .6 & 50 & 10 & 0.99897 \\
500 & .6 & 50 & 20 & 0.99892 \\
500 & .6 & 50 & 50 & 0.99888 \\
500 & .6 & 50 & 100 & 0.99887 \\
500 & .6 & 50 & 1000 & 0.99886\end{array}
$
\end{table}

We see in the table above that the relative efficiency of the two estimators is very close to one, even for small values of $k$. In the following section we will use the more efficient estimator of $N$, with $k_1=k_2=k$.

\section{Two-Stage and Sequential Estimators with prescribed Precision}
For precision of estimation, we determine the sample size so that $2(Var({\overline{N}}_{k}))^{1/2}$  will be smaller than $\gamma N$. This interval is called "Proportional Closeness Interval of Prescribed Precision" $1 -\alpha$. Here $0 <\gamma  <1.$  If the parameters $N$ and $p$ were known, we would need $K^{0}$ observations, which is the solution $k$ of the equation
\begin{equation}
\label{eq:Q}
4q(Np +m +q/k) =Nkmp\gamma ^{2}.
\end{equation}

Define $\xi =(1 -p)/k$ then Equation \eqref{eq:Q} becomes the quadratic equation

\begin{equation}
\label{eq:QA}
\xi^{2} +(Np+m)\xi-Nmp\gamma^2/4 =0.
\end{equation}

It follows that
\begin{align}
\label{eq:K}
&
\xi=\frac{(Np+m)\left(\left(1+\frac{Nmp\gamma^2}{(Np+m)^2}\right)^{1/2}-1\right)}{2}\nonumber,\\
&
K^{0} =k =(1 -p)/\xi.
\end{align}

Thus, if $N$ and $p$ were known, we would need, for the case of $N=500, p=0.6, m=10$ and $\gamma  =0.01$, $K^{0} =1654$ observations. However, since $N$ and $p$ are unknown parameters, we could take first a pilot sample of $K_{1}$ values of $X$ and $T$, estimate the parameters $N$ and $p$, and then substitute these estimates in the function \eqref{eq:K}.

\subsection{Two-Stage Sampling}

We apply the following two-stage sampling  procedure.
\bigskip

\noindent
{\bf{Stage 1}}:
Take random samples of size $K_{1}$ from the binomial $X$ trials and independently from the Pascal $T$ trials, and compute the sample means ${\overline{X}}_{K_{1}}$ and ${\overline{T}}_{K_{1}}$.
In equation \eqref{eq:K} substitute $p$ with $m/{\overline{T}}_{K_{1}}$ and $N$ with ${\overline{X}}_{K_{1}}{\overline{T}}_{K_{1}}/m$; and obtain a formula for the desired sample size in the second stage, i.e.
\begin{equation}
\label{eq:TS}
K_{TS}(K_1)=\Big\lfloor\frac{2({\overline{T}}_{K_{1}}-m)}{{\overline{T}}_{K_{1}}\left({\overline{X}}_{K_{1}}+m\right)
\left(
\left(1+\frac{{\overline{X}}_{K_{1}}m\gamma^2}{\left({\overline{X}}_{K_{1}}+m\right)^2}\right)^{1/2}
-1\right)
}
\Big\rfloor+1.
\end{equation}

If $K_{1} \geq K_{TS}(K_1)$  stop sampling and apply the available estimates of $N$ and $p$. On the other hand, if $K_{TS} >K_{1}$ go to Stage 2.
\bigskip

\noindent
{\bf{Stage 2}}:
Sample additional $K\ast=K_{TS} -K_{1}$ values of $X$  and $T$ independently. The estimators of $N$ and $p$  are $\overline{X}_{TS}\overline{T}_{TS}/m$  and $m/\overline{T}_{TS} $, respectively, where $K_{TS}=\max\left\{K_1, K_{TS}(K_1)\right\}$.
\medskip


In the following we obtain an asymptotic approximation for the expected values of $K_{TS}(K_1)$, for large values of $K_1$.
Let $${\displaystyle
\overline{Y}_{K_{1}}=\overline{T}_{K_{1}}-m,\,\, A\left(\overline{Y}_{K_{1}}\right)=\frac{2\overline{Y}_{K_{1}}}{\overline{Y}_{K_{1}}+m}
}.$$
In addition, let
$$B\left(\overline{X}_{K_{1}}\right)=
\left({\overline{X}}_{K_{1}}+m\right)
\left(
\left(1+\frac{{\overline{X}}_{K_{1}}m\gamma^2}{\left({\overline{X}}_{K_{1}}+m\right)^2}\right)^{1/2}
-1\right).$$
Therefore,
\begin{equation}
\label{eq:Zeut}
K_{TS}(K_1)=\Big\lfloor A\left(\overline{Y}_{K_{1}}\right) \frac{1}{B\left(\overline{X}_{K_{1}}\right)}  \Big\rfloor+1.
\end{equation}
Accordingly,
\begin{equation}
\label{eq:Ex}
E\left(K_{TS}(K_1)\right)=E\left(A\left(\overline{Y}_{K_{1}}\right)\right)E\left(\frac{1}{B\left(\overline{X}_{K_{1}}\right)}\right).
\end{equation}

By CLT, the asymptotic distribution of $\overline{Y}_{K_{1}}$ and $\overline{X}_{K_{1}}$ are normal with
$$E\left(\overline{Y}_{K_{1}}\right)=\frac{mq}{p},\,\, Var\left(\overline{Y}_{K_{1}}\right)=\frac{mq}{p^2K_1},\,\,
E\left(\overline{X}_{K_{1}}\right)=Np,\,\, Var\left(\overline{X}_{K_{1}}\right)=\frac{Npq}{K_1}.$$

By numerical integration in R we obtain, for $N=500,\, p=0.6,\, m=10,\, K_1=100,\, \gamma=0.1$\,\, $E\left(K_{TS}(100)\right)\cong
1652.393$. Remarkably, this value almost equals to $K^{0} =1654$.

In this similar manner, by using the second moments of \eqref{eq:Zeut} and \eqref{eq:Ex}
we found numerically the standard deviation under the 2-stage procedure as 49.6425.

We consider the following confidence interval for $N$,
\begin{equation}
\left[(1-\gamma) \overline{N}_{K_{TS}}, (1+\gamma) \overline{N}_{K_{TS}} \right],
\end{equation}

where ${\displaystyle \overline{N}_{K_{TS}}=\frac{(\overline{Y}_{K_{TS}}+m)\overline{X}_{K_{TS}}}{m}}$.
\smallskip

\noindent
The coverage probability, $CP$, of this interval is
\begin{align}
&
CP=P\left\{(1-\gamma) \overline{N}_{K_{TS}}\leq N \leq (1+\gamma) \overline{N}_{K_{TS}}\right\}=
P\left\{\frac{Nm}{(1+\gamma)\overline{X}_{K_{TS}}}-m\leq \overline{Y}_{K_{TS}}\leq \frac{Nm}{(1-\gamma)\overline{X}_{K_{TS}}}-m\right\}.
\end{align}

In the above example we evaluate CP by using normal approximation of
${\displaystyle \overline{Y}_{K_{TS}}\approx N\left(\frac{mq}{p}, \frac{mq}{p^2 K_{TS}}\right)}$
and ${\displaystyle \overline{X}_{K_{TS}}\approx N\left(NP, \frac{Npq}{K_{TS}}\right)}$,
substituting 1653 for $K_{TS}$. Numerically integrating with respect to density of ${\displaystyle \overline{X}_{K_{TS}}}$
over the range $[295, 305]$, we obtained $CP\cong 0.9544$.

\subsection{Sequential Sampling}

In the sequential procedure we start as before with a pilot sample of size $K_{1}$ and compute the critical value $K_{TS}(K_1) $. If $K_{1} >K_{TS}(K_1)$ we stop sampling, otherwise we continue sampling from $X$ and $T$ one by one
until stopping rule \eqref{eq:SE} is satisfied.

\begin{equation}
\label{eq:SE}
K_{SE}(K_1) =\inf \left\{k >K_{1} :k >K_{TS}(k)
\right\}.
\end{equation}

The final sample size in the sequential procedure is
$$K_{SE}=\max\left\{K_1, K_{SE}(K_1) \right\}.$$

In order to compare the efficiency of the sequential procedure relative to the two--stage procedure, we sample 1000 replicas of the sequential procedure with the R-function SeqK2 and a sample of 1000 replicas of the two--stage sample with the function TSK (Appendix \ref{se:A}). The mean, std, and (0.025,0.975) quantiles of the number of the observations are presented in the following table.

\begin{table}[H]
\caption {Simulated two--stage and sequential values of $K$ for $N=500,\,p=0.6,\, m=10,\,\gamma=0.01$}

$\begin{array}{cccc}Sample & Mean & std & (Q_{.025} ,Q_{.975}) \\
2 -stage   & 1653.04 & 49.478 & 1555, 1749 \\
Sequential & 1653.42 & 12.107 & 1630, 1677\end{array}$
\end{table}

Recall that $K^{0} =1654$. We see that the two sampling types yield close mean values of $K$. The std of the sequential is significantly smaller. As a result, $95\%$ of the sequential results are closer to the nominal value than those of the 2--stage. The sequential procedure is much more efficient than the 2--stage procedure. In order to demonstrate the characteristics of the sequential procedure, we present in the following table 1000 simulations of the properties of sequential procedures.

\begin{table}[H]
\caption {Sample statistics of 1000 simulations of the sequential procedure}
$
\begin{array}{llllllllll}
N & p & m & K_1 & \gamma & mean(K_{SE})  & mean({\overline{N}_{K_{SE}}}) & mean(\gamma{\overline{N}_{K_{SE}}})  \\
500 & .6 & 10 & 100 & 0.01    & 1653   & 499.98  & 4.999 \\
500 & .6 & 20 & 100 & 0.01    & 854    & 500.01  & 5.000\\
500 & .6 & 10 & 100 & 0.05    & 100    & 499.80  & 24.990 \\
500 & .6 & 20 & 100 &0.05     & 100    & 500.17  & 25.009 \\
100 & .4 & 10 & 100 &0.01     & 3001   & 100.02  & 1.000 \\
100 & .4 & 20 & 100 &0.01     & 1801   & 99.98   & 1.000 \\
100 & .4 & 10 & 100 &0.05     & 121    & 99.93   & 4.996 \\
100 & .4 & 20 & 100 &0.05     & 100    & 99.96   & 4.998\\
\end{array}
$
\end{table}

In this table we clearly see the effect of increasing $m$ on $K$.

\section*{Acknowledgement}
We would like to thank the Editor, Associate Editor, and three referees for the insightful and helpful comments that led to improvements in the paper.
The research of YM was supported by grant no. 2020063 from the United States--Israel Binational Science
Foundation (BSF), Jerusalem, Israel.

\section*{Appendix}
\appendix
\section{}
\label{se:A}

R functions for Section 4.\\

$
> K0\\
function(N,p,m,g)\{
\\
A<-(N*p+m)\\
B<-N*m*p*g^2\\
csi<-A*(sqrt(1+B/A^2)-1)/2\\
k<-(1-p)/csi\\
out<-floor(k)+1\\
out\\
\}
\\
> TSK
\\
\\
\noindent
function(N,p,m,k1,g,Ns)\{
\\
resTS<-c(1:Ns)\\
resN<-c(1:Ns)\\
resp<-c(1:Ns)\\
for(j\,\,\, \text{in}\,\,\, 1:Ns)\{
\\
k<-k1\\
XM<-sum(rbinom(k,N,p))/k\\
S<-0\\
for(i\,\,\, \text{in} \,\,\,1:k)\{
\\
S<-S+simPascal(m,p)\}
\\
TM<-S/k\\
KS<-KSE(XM,TM,m,g)\\
resTS[j]<-KS\\
XM2<-sum(rbinom(KS,N,p))/KS\\
Tk<-0\\
for(i\,\,\, \text{in}\,\,\,\, 1:KS)\{
\\
Tk<-Tk+simPascal(m,p)\}
\\
TM2<-Tk/KS\\
Ph<-m/TM2\\
Nh<-XM2*TM2/m\\
resN[j]<-Nh\\
resp[j]<-Ph\}
\\
MKS<-mean(resTS)\\
SKS<-std(resTS)\\
MN<-mean(resN)\\
Mp<-mean(resp)\\
out<-list(MKS,SKS,MN,Mp)\\
out\\
\}
\\
\\
\noindent
>> SeqK2\\
function(N,p,m,k1,g)\{\\
k<-k1\\
Xk<-sum(rbinom(k1,N,p))\\
XM<-Xk/k\\
Tk<-0\\
for(i\,\,\,\, \text{in}\,\,\, 1:k1)\{
\\
Tk<-Tk+simPascal(m,p)\}
\\
TM<-Tk/k\\
CR<-KSE(XM,TM,m,g)\\
repeat\{
\\
if(k>CR)\\
break\\
k<-k+1\\
Xk<-Xk+rbinom(1,N,p)\\
XM<-Xk/k\\
Tk<-Tk+simPascal(m,p)\\
TM<-Tk/k\\
CR<-KSE(XM,TM,m,g)\}
\\
Ph<-m/TM\\
Nh<-XM*TM/m\\
KS<-k\\
dlt<-Nh*g\\
out<-KS\\
out
\}
$

\end{document}